\documentclass{Interspeech}
\usepackage{makecell}
\usepackage{adjustbox}
\usepackage{bm}
\usepackage{amsbsy}
\usepackage{multirow}
\usepackage{algorithm} 
\usepackage{algpseudocode} 

\interspeechcameraready

\title{Vo-Ve: An Explainable Voice-Vector for Speaker Identity Evaluation\thanks{This paper has been accepted to Interspeech 2025.}}

\author[affiliation={1,2}]{Jaejun}{Lee}
\author[affiliation={1,2,3}]{Kyogu}{Lee}

\affiliation{}{Music and Audio Research Group (MARG)}{}
\affiliation{}{Department of Intelligence and Information}{}\affiliation{Artificial Intelligence Institute}{Seoul National University}{Republic of Korea}
\email{jjlee0721@snu.ac.kr, kglee@snu.ac.kr}

\usepackage{comment}

\begin{document}

\maketitle

% \noindent\textit{This paper has been accepted to Interspeech 2025.}
\begin{abstract}
In this paper, we propose Vo-Ve, a novel voice-vector embedding that captures speaker identity. Unlike conventional speaker embeddings, Vo-Ve is explainable, as it contains the probabilities of explicit voice attribute classes. Through extensive analysis, we demonstrate that Vo-Ve not only evaluates speaker similarity competitively with conventional techniques but also provides an interpretable explanation in terms of voice attributes. We strongly believe that Vo-Ve can enhance evaluation schemes across various speech tasks due to its high-level explainability.
\end{abstract}

\section{Introduction}
Recent advancements in human speech technologies, such as recognition and synthesis, focus not only on content but also on effectively capturing and imprinting the intended speaker identity.

Capturing speaker identity is crucial for tasks like speaker recognition~\cite{huh2024vox}. Additionally, multi-speaker settings have now become standard in speech synthesis tasks such as text-to-speech (TTS)~\cite{casanova2022yourtts, wang2023neural, le2024voicebox}, requiring the synthesized speech to resemble the target speaker’s voice. Voice conversion (VC), in particular, is a representative task that explicitly focuses on speaker identity.

Despite the significant progress in speech synthesis techniques, evaluating speaker identity remains a challenge. Most works in VC assess the ability to preserve speaker information by measuring the cosine similarity of speaker embeddings, which are typically extracted from a pretrained speaker verification model~\cite{desplanques2020ecapa, chen2022wavlm, resemblyzer}. While similarity values can demonstrate relative performance between the synthesized outputs of different models, they do not clarify what the absolute similarity value signifies. Moreover, conventional embedding similarity values fail to provide insights into which specific voice attributes contribute to similarity or dissimilarity.

In this research, we propose Vo-Ve, a novel voice vector for speaker identity evaluation, representing the probabilities of explicit voice attributes. \mbox{Vo-Ve} is derived from the outputs of a multi-label classification task, yet it proves to be a powerful tool. First, we explain how Vo-Ve is generated and demonstrate its ability to perform competitively with conventional speaker similarity evaluation techniques. More importantly, through experiments on practical applications, we show that Vo-Ve values are interpretable in terms of explicit voice attributes.

To contextualize, our contributions are as follows:
\begin{itemize}
\item We propose Vo-Ve, a novel explainable voice vector for evaluating speaker information.
\item We assess its effectiveness as a speaker embedding by comparing it with conventional speaker embedding similarity measures.
\item Through a newly proposed evaluation framework for practical applications, we demonstrate that Vo-Ve not only enables relative similarity assessments but also explains which attributes contribute to the results and to what extent.
\end{itemize}

\noindent
The implementation code for Vo-Ve is available on, \url{https://github.com/jaejunL/vove}.

\begin{table}[t!]
\caption{Voice attribute classes in Vo-Ve}
\centering
\begin{adjustbox}{width=0.32\textwidth}
\begin{tabular}{ |c|c|c|c| } 
 \hline
 adult-like & gender-neutral & modest & sincere\\ 
 bright & halting & muffled & soft \\ 
 calm & hard & nasal & strict \\ 
 clear & intellectual & old & sweet \\ 
 cool & intense & powerful & tensed \\ 
 cute & kind & raspy & thick \\ 
 dark & light & reassuring & thin \\ 
 elegant & lively & refreshing & unique \\ 
 feminine & masculine & relaxed & weak \\ 
 fluent & mature & sexy & wild \\ 
 friendly & middle-aged & sharp & young \\  
 \hline
\end{tabular}
\end{adjustbox}
\vspace{-4mm}
\label{table:label}
\end{table}

\section{Related works}

\subsection{Conventional evaluation of speaker identity}
Evaluating speaker identity in speech is crucial, especially in multi-speaker synthesis frameworks, where preserving the voice of the target speaker is essential. Voice conversion~(VC), one of the key synthesis criteria, assesses its effectiveness by comparing the cosine similarity of speaker embeddings.

The latest VC research has employed various speaker embeddings for evaluation. Some studies~\cite{yang2024rw, gengembre2024disentangling, tanaka2024prvae} utilize ECAPA-TDNN~\cite{desplanques2020ecapa}, while others~\cite{baade2024neural, gusev2024improvement, igarashi2024noise} adopt WavLM-TDNN~\cite{chen2022wavlm}. Another commonly used~\cite{um2024utilizing, lee2024hear, guo2024xe} embedding is Resemblyzer~\cite{resemblyzer}. All three speaker embeddings are intermediate representations extracted from pretrained speaker verification models, leveraging different methods such as classical generalized end-to-end~(GE2E)~\cite{wan2018generalized} loss~\cite{resemblyzer}, statistical pooling~\cite{desplanques2020ecapa}, or self-supervised learning~(SSL) representations~\cite{chen2022wavlm}.

While the similarities of these speaker embeddings have been standard, key challenges and open questions persist, warranting further investigation. Conventional speaker embeddings allow for relative comparisons, meaning they can indicate which model performs better based on similarity value. However, they fail to explain which specific voice attributes contirbute to the similarity or dissimilarity, as well as the extent to which each attribute influences the results. In other words, conventional speaker embedding lack interpretability.

\subsection{High-level voice attributes}
The need for generating speech with diverse speaking style and speaker identities has evolved into a significant research topic. Prompt-based TTS~\cite{zhou2022content, guo2023prompttts, zhang2023promptspeaker, leng2023prompttts} is one of the primary approaches, utilizing text descriptions not only for the content but also to control the speech style. However, most existing research is limited to low-level voice style attributes such as `\textit{loud voice}', `\textit{high pitch}', or `\textit{slow speaking}', meaning that it lacks full interpretability.

Recently, datasets incorporating high-level voice attributes have been introduced, containing descriptive categories such as `\textit{cute}', `\textit{elegant}', `\textit{muffled}', and `\textit{tensed}'. Shimizu~\textit{et al.}~\cite{shimizu2024prompttts++} constructed a high-level voice attribute dataset for 404 speakers. Kawamura~\textit{et al.}~\cite{kawamura2024libritts} further expanded this dataset to 2,443 speakers, introducing it as LibriTTS-P.

\section{Vo-Ve: An Explainable Voice-Vector} \label{sec:vove}
Vo-Ve is a vector representation ($v$) where each dimension corresponds to a specific voice attribute class, and its value represents the degree (0$\sim$1) to which that attribute is present. It is the sigmoid-activated output of a multi-label classification network trained on a voice attribute dataset.

\begin{algorithm}[t!]
\caption{Voice attribute class to ground-truth label $y$}
\begin{algorithmic}[1]
\label{algori}
\Require Annotated labels for 44 voice attributes, each with intensity levels: \textit{very}, \textit{normal}, \textit{slightly}, \textit{none}
% \State Weights $w$: $\textit{very}=1.5$, $\textit{normal}=1.25$, $\textit{slightly}=0.5$
\State Weights $w$: $\textit{very}=1.5$, $\textit{normal}=1.25$, $\textit{slightly}=0.5$, $\textit{none}=0$
\For{each speaker}
    \For{each attribute $i = 1$ to $44$}
        \State Retrieve three annotator labels $(l_1, l_2, l_3)$ for attribute $i$
        \State Compute weighted sum: $s = w(l_1) + w(l_2) + w(l_3)$
        \State Compute averaged score: $y_i = s / 3$
        \State Clip $y_i$ to range $[0,1]$
    \EndFor
\EndFor
\end{algorithmic}
\end{algorithm}

\subsection{Voice attribute dataset} \label{sec:dataset}
To train the multi-label classification so that the output $v$ has the meaningful dimension, we used the LibriTTS-P~\cite{kawamura2024libritts}, the largest recently published voice attribute dataset. It containing 2,443 speakers and the voice attribute classes are listed in Table~\ref{table:label}.
LibriTTS-P is built upon speech data from \mbox{LibriTTS-R}~\cite{koizumi2023libritts} which is a widely used, improved sound quality version of \mbox{LibriTTS}~\cite{zen2019libritts}.
Three professional annotators labeled each speaker in the dataset with 44 voice attribute classes, each assigned an intensity level of \{\textit{very}, \textit{normal}, and \textit{slightly}\}. To convert the voice attribute class into a ground-truth label $y$ for classification network, we applied the process described in Algorithm~1.
The weights assigned to each intensity level were carefully chosen to ensure that stronger indications of an attribute contribute more significantly to the final score, while weaker indications have proportionally less impact. The assigned weights were determined such that a hard degree of 1 is granted for the $i$-th attribute of the ground-truth label ($y_i$) if:
\begin{enumerate}
\item At least two annotators labeled the $i$-th attribute as ``\textit{very}", or
\item All three annotators provided a label combination with at least the intensive with \{\textit{normal}, \textit{normal}, \textit{slightly}\}.
\end{enumerate}
In all other cases, $y_i$ is assigned a soft degree value.

\subsection{Multi-label classification}
To ensure the reliability of Vo-Ve for unseen speakers, we trained a multi-label classification network using speech data from LibriTTS-R, with the ground-truth label $y$, as described in Section~\ref{sec:dataset}. The output of the multi-label classification, $v$, is a 44-dimensional soft vector, where each dimension corresponds to an explicit voice attribute class (Table~\ref{table:label}), and its value represents the probability degree of that class. The classification network is based on the ECAPA-TDNN~\cite{desplanques2020ecapa} architecture with a fully connected classification layer. Additionally, we incorporate a speaker verification layer after the classification layer's output to enhance the model's inter-speaker discriminative ability. Finally, the total loss function $\mathcal{L}_{total}$ is defined as follows,
\vspace{-2mm}
\begin{equation}
v = \sigma(f(x))
\label{eq:v}
\end{equation}
\vspace{-4mm}
\begin{equation}
\mathcal{L}_{total}=\mathcal{L}_{\mathit{BCE}}(v,y)+\mathcal{L}_{\mathit{CE}}(\phi(g(f(x))),s)
\label{eq:loss}
\end{equation}
where $\sigma(f(x))$ represents a multi-label classification network, while $\mathcal{L}_{\mathit{BCE}}$ and $\mathcal{L}_{\mathit{CE}}$ refer to binary cross-entropy (BCE) loss and cross-entropy (CE) loss, respectively. The input $x$ is the log-Mel spectrogram of the speech, and $s$ denotes the speaker index. The function $g$ consists of a ReLU activation, followed by a fully-connected layer with batch normalization. Additionally, $\sigma$ and $\phi$ represent the sigmoid and softmax activation functions, respectively.

We trained the classification model using the predefined training split in LibriTTS-R, utilizing only the speech data for which a corresponding speaker exists in LibriTTS-P. The model was trained for 30~epochs based on validation loss, using AdamW~\cite{loshchilov2017decoupled} with a learning rate of 0.0001.

\begin{table}[t!]
\caption{Multi-label classification performance}
\centering
\begin{adjustbox}{width=0.48\textwidth}
\begin{tabular}{c|c|c|c}
\hline\hline
\makecell{Threshold\\$\tau$} & \makecell{Precision\\score} & \makecell{Recall\\score} & \makecell{$F_1$\\score}\\
\hline
0.1 & $\pmb{0.9996 \pm 0.0039}$ & $0.6274 \pm 0.0597$ & $0.7692 \pm 0.0470$ \\ 
0.2 & $0.9861 \pm 0.0214$ & $0.6833 \pm 0.0747$ & $0.8047 \pm0.0549$ \\ 
0.3 & $0.9488 \pm 0.0429$ & $0.7220 \pm 0.0777$ & \pmb{$0.8176 \pm0.0582$} \\
0.4 & $0.8581 \pm 0.0649$ & \pmb{$0.7515 \pm 0.0813$} & $0.7988 \pm 0.0632$ \\
0.5 & $0.7596 \pm 0.1212$ & $0.3876 \pm 0.0806$ & $0.5094 \pm 0.0905$ \\ \hline
\hline
\end{tabular}
\end{adjustbox}
\label{table:result1}
\vspace{-2mm}
\end{table}

\subsection{Classification performance}
To evaluate the performance of multi-label classification, we measured the precision, recall, and $F_1$ scores. Since the ground-truth label~$y$ is a soft vector, we applied a threshold~($\tau$) to convert it into a hard label. The results, shown in Table~\ref{table:result1}, indicate that precision decreases as $\tau$ increases, meaning fewer labels are detected. The recall score is highest at $\tau=0.4$, while the $F_1$ score peaks at $\tau=0.3$. These findings demonstrate that our classification network performs well, particularly when $\tau<0.5$.

\section{Leveraging Vo-Ve: From Evaluation to Practical Applications}
In this section, we evaluate the capability of Vo-Ve on unseen datasets. First, we assess speaker embedding similarity, a conventional method for evaluating speaker identity, by comparing Vo-Ve with conventional speaker embeddings~\cite{desplanques2020ecapa, chen2022wavlm, resemblyzer} and demonstrating its comparable performance.
More importantly, we introduce two evaluations focusing on inter-speaker and intra-speaker interpretability, highlighting Vo-Ve’s practical applications in a novel way.
To apply Vo-Ve to unseen datasets, we used the same model from Section~\ref{sec:vove}, but trained it on the entire dataset without a split, applying early stopping at 10~epochs based on training loss. The following experiments are based on the output vector $v$ from the multi-label classification network, which represents the probability degrees of explicit voice attribute classes.

\begin{table}[t!]
\caption{Speaker embedding similarity evaluation results. ECAPA refers to ECAPA\_TDNN~\cite{desplanques2020ecapa}, WavLM refers to WavLM\_TDNN~\cite{chen2022wavlm}, Resem refers to Resemblyzer~\cite{resemblyzer}, and Vo-Ve represents the proposed method.}
\centering
\begin{adjustbox}{width=0.48\textwidth}
\begin{tabular}{cccccc}
\hline\hline
\multirow{2}{*}{Models} & \multirow{2}{*}{Homogeneity} & \multirow{2}{*}{Diversity} & \multicolumn{3}{c}{Top-$k$ accuracy $\%$ ($\uparrow$)}\\ \cmidrule(lr){4-6}
& ($\uparrow$) & ($\downarrow$) & $k=1$ & $k=5$ & $k=10$\\
\hline
ECAPA & $0.6243$ & $0.1164$ & $97.56$ & $99.72$ & $99.90$\\
WavLM & $0.9081$ & $0.6346$ & $53.00$ & $79.67$ & $88.51$\\ 
Resem & $0.7995$ & $0.5477$ & $78.10$ & $94.06$ & $97.31$\\ 
Vo-Ve & $0.9862$ & $0.9263$ & $63.29$ & $87.31$ & $93.63$\\ 
\hline\hline
\end{tabular}
\end{adjustbox}
\label{table:result2}
\vspace{-2mm}
\end{table}

\subsection{Speaker embedding similarity evaluation} \label{subsec:sim}
To evaluate Vo-Ve's ability in speaker embedding similarity on an unseen dataset, we chose VCTK dataset~\cite{yamagishi2019cstr}, which contains 110 speakers, each with approximately 400 speech sentences. We compared Vo-Ve with conventional pretrained speaker embeddings, ECAPA\_TDNN~\cite{desplanques2020ecapa} with its implementation~\footnote{https://huggingface.co/speechbrain/spkrec-ecapa-voxceleb}, WavLM\_TDNN~\cite{chen2022wavlm} with its implementation~\footnote{https://huggingface.co/microsoft/wavlm-base-plus-sv}, and Resemblyzer~\cite{resemblyzer} with its implementation~\footnote{https://github.com/resemble-ai/Resemblyzer}.
None of the four models were trained on the VCTK dataset, ensuring an unseen data setting. Strictly speaking, the training datasets for each model differ; however, each follows a widely adopted implementation setting, making the evaluation meaningful. Note that our goal is not to achieve the best performance in similarity evaluation but rather to compare Vo-Ve with conventional standard metrics.

The results are presented in Table~\ref{table:result2}. For all evaluation metrics, we performed a paired t-test for each pair of models, and all results showed statistical significance with $p<0.01$.
\begin{itemize}
    \item Homogeneity is defined as the cosine similarity of speaker embeddings within the same speaker, where higher similarity values are expected. We randomly selected 100 speech samples per speaker and measured the similarity of all possible pairs, excluding self-pairs. The average similarity over all speakers was reported.
    \item Diversity refers to the cosine similarity of speaker embeddings from different speakers. Unlike homogeneity, in this case, the model aims to capture distinct speaker identities, so the similarity values are expected to be low. We randomly selected one speech sample per speaker and measured the similarity of all possible pairs among them. This process was repeated 100 times, and the average similarity was reported.
    \item Top-$k$ accuracy measures whether the speaker embedding of the corresponding ground-truth speaker appears within the top $k$ most similar embeddings to the query embedding. We randomly selected one speech sample per speaker and measured its similarity with the query. The query sample was randomly selected from each speaker, and the corresponding ground-truth embedding was taken from a different speech sample of the same speaker. This process was repeated 100 times, and the average accuracy was reported.
\end{itemize}

In the homogeneity metric, the proposed Vo-Ve exhibited the highest similarity. However, in the diversity metric, Vo-Ve demonstrated weaker discriminative ability. This limitation is likely influenced by its significantly smaller dimensionality (44 dimensions) compared to the other models (192 for ECAPA\_TDNN, 512 for WavLM\_TDNN, and 256 for Resemblyzer), which may result in reduced expressiveness due to its representational space.
For top-$k$ accuracy, although ECAPA\_TDNN achieved the best performance across all $k$ value, Vo-Ve performed comparably to other models and notably outperformed WavLM\_TDNN.
While the results of the diversity metric may suggest that Vo-Ve lacks inter-speaker discriminative ability, the results of the top-$k$ accuracy metric, a more critical measure of inter-speaker discrimination, indicate that Vo-Ve possesses proper discriminative ability comparable to other models.
Thess results are particularly significant because, unlike the other models, Vo-Ve represents only the probability of explicit voice attribute classes, yet it still performs competitively with conventional speaker embeddings. This suggests that Vo-Ve offers interpretability while maintaining comparable speaker embedding performance with minimal trade-offs.

\subsection{Interpretable application of Vo-Ve}
The greatest advantage of using Vo-Ve is that all values in the vector are interpretable, as each dimension corresponds to an explicit voice attribute, and its value represents the degree to which that attribute is present. To verify this capability, we conduct two experiments focusing on inter-speaker and intra-speaker interpretability in a novel way that has not been previously explored.

\begin{table}[t!]
\caption{Inter-speaker ABX subjective test results}
\centering
\begin{adjustbox}{width=0.27\textwidth}
\begin{tabular}{c|c}
\hline\hline
\makecell{Evaluation set} & \makecell{accuracy (\%)}\\
\hline
\makecell{\textbf{Inter-speaker}\\\textbf{dissimilar pair set}} & ${53.95}$\\ 
\hline
\makecell{\textbf{Inter-speaker}\\\textbf{similar pair set}} & ${49.74}$\\ 
\hline\hline
\end{tabular}
\end{adjustbox}
\label{table:result3}
\vspace{-2mm}
\end{table}

\subsubsection{Inter-speaker interpretability} \label{subsec:inter}
For inter-speaker interpretability, we used the VCTK dataset, as described in Section~\ref{subsec:sim}. The VCTK dataset is built upon predefined scripts, meaning that speech samples from different speakers contain identical text content. To minimize the effect of content variability, we randomly selected two speech samples from different speakers with the same text and compared their Vo-Ve representations ($v$). We configured two types of evaluation sets:
\begin{enumerate}
    \item \textbf{Inter-speaker dissimilar pair set} : This set consists of speech pairs with a noticeable difference in $v$ for a specific attribute dimension. Here, we define a noticeable difference as greater than 0.3. This means that the two speech samples exhibit a clear discriminative point in a specific voice attribute. For example, if Speech~A has a value of 0.8 in the 3rd dimension (corresponding to the class \textit{calm} as shown in Table~\ref{table:label}), while Speech~B has a value of 0.4 in the same dimension, we include this pair in the evaluation set with the label \textit{calm}.
    \item \textbf{Inter-speaker similar pair set} : This set consists of pairs of speech samples with a corresponding voice attribute label, where the difference between the two speech samples in a given voice attribute dimension is less than 0.1.
\end{enumerate}

To validate our approach, we conducted an ABX subjective test using Amazon Mechanical Turk (MTurk). Participants were asked to determine which speech sample better matched the given voice attribute label. If the chosen speech sample had a higher $v_i$ value in the corresponding $i$-th label dimension than the other sample, it would indicate that differences in $v_i$ between speakers are interpretable.

Each evaluation set consisted of 100 speech pairs with randomly chosen labels, and participants were assigned four pairs per evaluation set. To identify unreliable annotators, we included a fake sample pair that was clearly not human speech.
It is important to note that the speech pairs were selected under a controlled gender setting to ensure that distinctions are not solely based on obvious attributes such as \textit{feminine} or \textit{masculine}, making it a more challenging condition for Vo-Ve. Each participant was rewarded \$1.50 USD for evaluating nine pairs, and the total number of participants was 100.

The results of the averaged ABX test accuracy are presented in Table~\ref{table:result3}. The accuracy indicates the proportion of participants who selected the speech sample that Vo-Ve predicted to have a higher $v_i$ value for the given label compared to the other sample.
For the \textbf{Inter-speaker dissimilar pair set} results, despite the strict controlled gender setting, Vo-Ve performed significantly better than random chance. This suggests that Vo-Ve provides meaningful interpretability for specific voice attribute classes.
Results from the \textbf{Inter-speaker similar pair set} showed insignificant accuracy, indicating that when Vo-Ve predicted only a minimal difference in a given voice attribute, participants similarly found it difficult to distinguish between the speech samples. This implies that Vo-Ve not only offers interpretability in a discriminative manner but also assigns values that meaningfully reflect the extent of a given voice attribute class.

\begin{table}[t!]
\caption{Intra-speaker ABX subjective test results}
\centering
\begin{adjustbox}{width=0.27\textwidth}
\begin{tabular}{c|c}
\hline\hline
\makecell{Evaluation set} & \makecell{accuracy (\%)}\\
\hline
\makecell{\textbf{Intra-speaker}\\\textbf{dissimilar pair set}} & ${56.28}$\\ 
\hline
\makecell{\textbf{Intra-speaker}\\\textbf{similar pair set}} & ${49.18}$\\ 
\hline\hline
\end{tabular}
\end{adjustbox}
\label{table:result4}
\vspace{-2mm}
\end{table}

\subsubsection{Intra-speaker interpretability}

Unlike inter-speaker interpretability applications, we conducted an intra-speaker interpretability evaluation in the context of assessing speech synthesis systems. Specifically, we evaluated a voice conversion (VC) system, which aims to replicate the target speaker’s voice.
To make the evaluation more rigorous, we assumed a more challenging condition: face-based voice conversion, which uses only the target speaker's facial image—rather than their voice—to mimic the original speech. For this, we utilized the recently published benchmark, \mbox{HYFace}~\cite{lee2024hear} with its implementation~\footnote{https://github.com/jaejunL/HYFace/}. HYFace is built upon the LRS3 dataset~\cite{afouras2018lrs3}, which consists of TED talk video data.

We compare ground-truth (GT) speech with synthesized speech that contains the same content as the GT speech and is generated using the corresponding face image of the target speaker. Note that for each speaker, multiple face images are available. To minimize the effect of content variability, we generated speech using all available face images and selected the sample that achieved the best word error rate (WER) performance. For the WER metric, we used Whisper~\cite{radford2023robust}, ``\textit{medium.en}'' model. Similar to Section~\ref{subsec:inter}, we constructed two evaluation sets using the predefined test split of the LRS3 dataset:
\begin{enumerate}
    \item \textbf{Intra-speaker dissimilar pair set} : This set consists of speech pairs where the difference in a given attribute dimension is noticeable (greater than 0.3).
    \item \textbf{Intra-speaker similar pair set} : This set consists of speech pairs where the difference in a given attribute dimension is minimal (less than 0.1).
\end{enumerate}
The composition of the pairs is identical to the inter-speaker evaluation setting, except that in this evaluation, each pair consists of two speech samples from the same speaker—one being the GT speech and the other synthesized using HYFace. A similar ABX subjective test was conducted, and the results are presented in Table~\ref{table:result4}.

According to the results of the \textbf{Intra-speaker dissimilar pair set}, similar to the inter-speaker evaluation, Vo-Ve demonstrated the ability to assess speech in a specific voice attribute manner. What we emphasize here is that, by using Vo-Ve, a speech synthesis system can diagnose which attributes are more or less similar to the ground-truth speech. This presents a promising foundation for advancing various speech synthesis techniques. For the \mbox{\textbf{Intra-speaker similar pair set}}, the results show insignificant accuracy, aligning with the findings in Section~\ref{subsec:inter}. This confirms that Vo-Ve not only offers interpretability in a discriminative manner but also that its absolute values carry meaningful information regarding the extent of a given voice attribute.

\section{Discussion}
Through extensive analysis, we demonstrated that Vo-Ve provides strong interpretability, a key characteristic lacking in conventional speaker embedding techniques. To fairly evaluate its potential, we designed a network architecture similar to that of the comparison system. However, for practical deployment across a wide range of real-world data, robustness to diverse datasets is essential. Techniques such as data augmentation or pseudo-label-based semi-supervised learning could help enhance its adaptability. We are actively working on improving the robustness of Vo-Ve for broader applications.
Additionally, we observed intriguing patterns—when the global pitch of speech was gradually lowered, the value of the \textit{sexy} dimension increased, while the \textit{bright} dimension exhibited the opposite trend. These findings suggest that certain voice attributes may have underlying correlations with pitch variations. Further investigation of such relationships is an exciting direction for future work.

\section{Conclusion}
In this work, we introduced Vo-Ve, an explainable voice vector designed for evaluating speaker identity. Unlike conventional speaker embeddings, Vo-Ve provides explicit voice attribute-based interpretability while maintaining competitive performance in speaker similarity evaluations.
Through extensive experiments, we demonstrated that Vo-Ve not only enables interpretability in a discriminative voice attribute manner but also assigns meaningful values that reflect the degree of each attribute. We anticipate that ongoing advancements in speech synthesis techniques could greatly benefit from Vo-Ve.

\section{Acknowledgements}
This work was partly supported by Institute of Information \& communications Technology Planning \& Evaluation (IITP) grant funded by the Korea government(MSIT) [No. RS-2022-II220641, 50\%], [No. RS-2022-II220320, 2022-0-00320, 40\%], [No.RS-2021-II211343, Artificial Intelligence Graduate School Program (Seoul National University), 5\%], and [No.RS-2021-II212068, Artificial Intelligence Innovation Hub, 5\%].

\bibliographystyle{IEEEtran}
\bibliography{main}

\begin{thebibliography}{10}
\providecommand{\url}[1]{#1}
\csname url@samestyle\endcsname
\providecommand{\newblock}{\relax}
\providecommand{\bibinfo}[2]{#2}
\providecommand{\BIBentrySTDinterwordspacing}{\spaceskip=0pt\relax}
\providecommand{\BIBentryALTinterwordstretchfactor}{4}
\providecommand{\BIBentryALTinterwordspacing}{\spaceskip=\fontdimen2\font plus
\BIBentryALTinterwordstretchfactor\fontdimen3\font minus \fontdimen4\font\relax}
\providecommand{\BIBforeignlanguage}[2]{{%
\expandafter\ifx\csname l@#1\endcsname\relax
\typeout{** WARNING: IEEEtran.bst: No hyphenation pattern has been}%
\typeout{** loaded for the language `#1'. Using the pattern for}%
\typeout{** the default language instead.}%
\else
\language=\csname l@#1\endcsname
\fi
#2}}
\providecommand{\BIBdecl}{\relax}
\BIBdecl

\bibitem{huh2024vox}
J.~Huh, J.~S. Chung, A.~Nagrani, A.~Brown, J.-w. Jung, D.~Garcia-Romero, and A.~Zisserman, ``The vox celeb speaker recognition challenge: A retrospective,'' \emph{IEEE/ACM Transactions on Audio, Speech, and Language Processing}, 2024.

\bibitem{casanova2022yourtts}
E.~Casanova, J.~Weber, C.~D. Shulby, A.~C. Junior, E.~G{\"o}lge, and M.~A. Ponti, ``Yourtts: Towards zero-shot multi-speaker tts and zero-shot voice conversion for everyone,'' in \emph{International Conference on Machine Learning}.\hskip 1em plus 0.5em minus 0.4em\relax PMLR, 2022, pp. 2709--2720.

\bibitem{wang2023neural}
C.~Wang, S.~Chen, Y.~Wu, Z.~Zhang, L.~Zhou, S.~Liu, Z.~Chen, Y.~Liu, H.~Wang, J.~Li \emph{et~al.}, ``Neural codec language models are zero-shot text to speech synthesizers,'' \emph{arXiv preprint arXiv:2301.02111}, 2023.

\bibitem{le2024voicebox}
M.~Le, A.~Vyas, B.~Shi, B.~Karrer, L.~Sari, R.~Moritz, M.~Williamson, V.~Manohar, Y.~Adi, J.~Mahadeokar \emph{et~al.}, ``Voicebox: Text-guided multilingual universal speech generation at scale,'' \emph{Advances in neural information processing systems}, vol.~36, 2024.

\bibitem{desplanques2020ecapa}
B.~Desplanques, J.~Thienpondt, and K.~Demuynck, ``Ecapa-tdnn: Emphasized channel attention, propagation and aggregation in tdnn based speaker verification,'' \emph{arXiv preprint arXiv:2005.07143}, 2020.

\bibitem{chen2022wavlm}
S.~Chen, C.~Wang, Z.~Chen, Y.~Wu, S.~Liu, Z.~Chen, J.~Li, N.~Kanda, T.~Yoshioka, X.~Xiao \emph{et~al.}, ``Wavlm: Large-scale self-supervised pre-training for full stack speech processing,'' \emph{IEEE Journal of Selected Topics in Signal Processing}, vol.~16, no.~6, pp. 1505--1518, 2022.

\bibitem{resemblyzer}
``Resemblyzer,'' \url{https://github.com/resemble-ai/Resemblyzer}.

\bibitem{yang2024rw}
C.-Y. Yang, S.~G. Upadhyay, Y.-T. Wu, B.-H. Su, and C.-C. Lee, ``Rw-voiceshield: Raw waveform-based adversarial attack on one-shot voice conversion,'' in \emph{Proc. Interspeech 2024}, 2024, pp. 2730--2734.

\bibitem{gengembre2024disentangling}
N.~Gengembre, O.~Le~Blouch, and C.~Gendrot, ``Disentangling prosody and timbre embeddings via voice conversion,'' in \emph{Proc. Interspeech 2024}, 2024, pp. 2765--2769.

\bibitem{tanaka2024prvae}
K.~Tanaka, H.~Kameoka, T.~Kaneko, and Y.~Kondo, ``Prvae-vc2: Non-parallel voice conversion by distillation of speech representations,'' in \emph{Proc. Interspeech 2024}, 2024, pp. 4363--4367.

\bibitem{baade2024neural}
A.~Baade, P.~Peng, and D.~Harwath, ``Neural codec language models for disentangled and textless voice conversion,'' in \emph{Proc. Interspeech 2024}, 2024, pp. 182--186.

\bibitem{gusev2024improvement}
A.~Gusev and A.~Avdeeva, ``Improvement speaker similarity for zero-shot any-to-any voice conversion of whispered and regular speech,'' in \emph{Proc. Interspeech 2024}, 2024, pp. 2735--2739.

\bibitem{igarashi2024noise}
T.~Igarashi, Y.~Saito, K.~Seki, S.~Takamichi, R.~Yamamoto, K.~Tachibana, and H.~Saruwatari, ``Noise-robust voice conversion by conditional denoising training using latent variables of recording quality and environment,'' \emph{arXiv preprint arXiv:2406.07280}, 2024.

\bibitem{um2024utilizing}
J.~S. Um and H.~Kim, ``Utilizing adaptive global response normalization and cluster-based pseudo labels for zero-shot voice conversion,'' in \emph{Proc. Interspeech 2024}, 2024, pp. 2740--2744.

\bibitem{lee2024hear}
J.~Lee, Y.~Oh, I.~Hwang, and K.~Lee, ``Hear your face: Face-based voice conversion with f0 estimation,'' in \emph{Proc. Interspeech 2024}, 2024, pp. 4378--4382.

\bibitem{guo2024xe}
H.~Guo, C.~Liu, C.~T. Ishi, and H.~Ishiguro, ``Xe-speech: Joint training framework of non-autoregressive cross-lingual emotional text-to-speech and voice conversion,'' in \emph{Proc. Interspeech 2024}, 2024, pp. 4983--4987.

\bibitem{wan2018generalized}
L.~Wan, Q.~Wang, A.~Papir, and I.~L. Moreno, ``Generalized end-to-end loss for speaker verification,'' in \emph{2018 IEEE International Conference on Acoustics, Speech and Signal Processing (ICASSP)}.\hskip 1em plus 0.5em minus 0.4em\relax IEEE, 2018, pp. 4879--4883.

\bibitem{zhou2022content}
Y.~Zhou, C.~Song, X.~Li, L.~Zhang, Z.~Wu, Y.~Bian, D.~Su, and H.~Meng, ``Content-dependent fine-grained speaker embedding for zero-shot speaker adaptation in text-to-speech synthesis,'' \emph{arXiv preprint arXiv:2204.00990}, 2022.

\bibitem{guo2023prompttts}
Z.~Guo, Y.~Leng, Y.~Wu, S.~Zhao, and X.~Tan, ``Prompttts: Controllable text-to-speech with text descriptions,'' in \emph{ICASSP 2023-2023 IEEE International Conference on Acoustics, Speech and Signal Processing (ICASSP)}.\hskip 1em plus 0.5em minus 0.4em\relax IEEE, 2023, pp. 1--5.

\bibitem{zhang2023promptspeaker}
Y.~Zhang, G.~Liu, Y.~Lei, Y.~Chen, H.~Yin, L.~Xie, and Z.~Li, ``Promptspeaker: Speaker generation based on text descriptions,'' in \emph{2023 IEEE Automatic Speech Recognition and Understanding Workshop (ASRU)}.\hskip 1em plus 0.5em minus 0.4em\relax IEEE, 2023, pp. 1--7.

\bibitem{leng2023prompttts}
Y.~Leng, Z.~Guo, K.~Shen, X.~Tan, Z.~Ju, Y.~Liu, Y.~Liu, D.~Yang, L.~Zhang, K.~Song \emph{et~al.}, ``Prompttts 2: Describing and generating voices with text prompt,'' \emph{arXiv preprint arXiv:2309.02285}, 2023.

\bibitem{shimizu2024prompttts++}
R.~Shimizu, R.~Yamamoto, M.~Kawamura, Y.~Shirahata, H.~Doi, T.~Komatsu, and K.~Tachibana, ``Prompttts++: Controlling speaker identity in prompt-based text-to-speech using natural language descriptions,'' in \emph{ICASSP 2024-2024 IEEE International Conference on Acoustics, Speech and Signal Processing (ICASSP)}.\hskip 1em plus 0.5em minus 0.4em\relax IEEE, 2024, pp. 12\,672--12\,676.

\bibitem{kawamura2024libritts}
M.~Kawamura, R.~Yamamoto, Y.~Shirahata, T.~Hasumi, and K.~Tachibana, ``Libritts-p: A corpus with speaking style and speaker identity prompts for text-to-speech and style captioning,'' \emph{arXiv preprint arXiv:2406.07969}, 2024.

\bibitem{koizumi2023libritts}
Y.~Koizumi, H.~Zen, S.~Karita, Y.~Ding, K.~Yatabe, N.~Morioka, M.~Bacchiani, Y.~Zhang, W.~Han, and A.~Bapna, ``Libritts-r: A restored multi-speaker text-to-speech corpus,'' \emph{arXiv preprint arXiv:2305.18802}, 2023.

\bibitem{zen2019libritts}
H.~Zen, V.~Dang, R.~Clark, Y.~Zhang, R.~J. Weiss, Y.~Jia, Z.~Chen, and Y.~Wu, ``Libritts: A corpus derived from librispeech for text-to-speech,'' \emph{arXiv preprint arXiv:1904.02882}, 2019.

\bibitem{loshchilov2017decoupled}
I.~Loshchilov, ``Decoupled weight decay regularization,'' \emph{arXiv preprint arXiv:1711.05101}, 2017.

\bibitem{yamagishi2019cstr}
J.~Yamagishi, C.~Veaux, K.~MacDonald \emph{et~al.}, ``Cstr vctk corpus: English multi-speaker corpus for cstr voice cloning toolkit (version 0.92),'' \emph{University of Edinburgh. The Centre for Speech Technology Research (CSTR)}, pp. 271--350, 2019.

\bibitem{afouras2018lrs3}
T.~Afouras, J.~S. Chung, and A.~Zisserman, ``Lrs3-ted: a large-scale dataset for visual speech recognition,'' \emph{arXiv preprint arXiv:1809.00496}, 2018.

\bibitem{radford2023robust}
A.~Radford, J.~W. Kim, T.~Xu, G.~Brockman, C.~McLeavey, and I.~Sutskever, ``Robust speech recognition via large-scale weak supervision,'' in \emph{International conference on machine learning}.\hskip 1em plus 0.5em minus 0.4em\relax PMLR, 2023, pp. 28\,492--28\,518.

\end{thebibliography}

\end{document}